# Deep learning based subdivision approach for large scale macromolecules structure recovery from electron cryo tomograms

Min Xu[1]*, Xiaoqi Chai[2], Hariank Muthakana[3], Xiaodan Liang[4],
Ge Yang[2], Tzviya Zeev-Ben-Mordehai[5], and Eric Xing[4]

[1] Computational Biology Department, Carnegie Mellon University, Pittsburgh, 15213, USA.
[2] Biomedical Engineering Department, Carnegie Mellon University, Pittsburgh, 15213, USA.
[3] Computer Science Department, Carnegie Mellon University, Pittsburgh, 15213, USA.
[4] Machine Learning Department, Carnegie Mellon University, Pittsburgh, 15213, USA.
[5] Division of Structural Biology, Wellcome Trust Centre for Human Genetics, University of Oxford, Oxford, OX3 7BN, UK.

## Abstract

**Motivation:** Cellular Electron CryoTomography (CECT) enables 3D visualization of cellular organization at near-native state and in sub-molecular resolution, making it a powerful tool for analyzing structures of macromolecular complexes and their spatial organizations inside single cells. However, high degree of structural complexity together with practical imaging limitations make the systematic *de novo* discovery of structures within cells challenging. It would likely require averaging and classifying millions of subtomograms potentially containing hundreds of highly heterogeneous structural classes. Although it is no longer difficult to acquire CECT data containing such amount of subtomograms due to advances in data acquisition automation, existing computational approaches have very limited scalability or discrimination ability, making them incapable of processing such amount of data.

**Results:** To complement existing approaches, in this paper we propose a new approach for subdividing subtomograms into smaller but relatively homogeneous subsets. The structures in these subsets can then be separately recovered using existing computation intensive methods. Our approach is based on supervised structural feature extraction using deep learning, in combination with unsupervised clustering and reference-free classification. Our experiments show that, compared to existing unsupervised rotation invariant feature and pose-normalization based approaches, our new approach achieves significant improvements in both discrimination ability and scalability. More importantly, our new approach is able to discover new structural classes and recover structures that *do not exist* in training data.

## 1 Introduction

Cellular processes are governed by macromolecules. Knowledge of the structures and spatial organizations of macromolecules within a single cell is a prerequisite for our understanding of biological processes. Cellular Electron CryoTomography (CECT) [22, 34, 24] enables the 3D visualization of structures at close-to-native state and in sub-molecular resolution within single cells [44, 39, 27, 4]. Therefore, if we knew how to systematically mine structures in cryo cellular tomograms, we would gain the desired knowledge on macromolecules' native structures and organization in their cellular context [41].

Systematic recovery of macromolecules structures from cryo tomograms is a very difficult task for several reasons. First, the cellular environment is very crowded [8, 17] with macromolecules that typically adopt different conformations as part of their function. Moreover, one macromolecule interacting with several different macromolecules can dynamically form different complexes at every time point. Therefore, a cellular tomogram has very complex and highly heterogeneous structural content. Second, the sizes of macromolecular complexes are typically smaller than 30nm, which is only slightly larger than the image resolution (~4nm). Finally, there are inherent practical limitations in data acquisition, in the form of low signal to noise ratio (SNR) and missing wedge effects.

Given the above challenges, successful systematic analysis of the macromolecule structures in CECT data relies on processing large amount of particles [3], possibly at least millions of particles and hundreds of structural classes. Nowadays, new imaging technologies and advances in automation allow a research lab to obtain hundreds of tomograms within several days [38], potentially containing millions of particles represented by 3D subimages (aka subtomograms). However, existing computational approaches have very limited discrimination ability or scalability, making them generally incapable for systematic *de novo* structural discovery on these amounts of particles.

Early works of analyzing the macromolecular complexes in CECT data focused at locating instances of macromolecular complexes in cells through template search [e.g. 41, 6]. However, such approaches do not discover new structures. For the reconstruction of novel structures repeating within cryo tomograms [15], reference-free subtomogram averaging [10], classification [e.g. 5, 55, 13, 47, 9] and structural pattern mining [59] methods have been developed. These approaches are essentially unsupervised clustering approaches, and they do not rely on any training data containing subtomograms with structural class labels. However, the scalability of such approaches is very limited, due to computationally intensive steps such as subtomogram alignment or integration over the six-dimensional rigid transformation space.

---

*Corresponding author

To complement the above approaches, rotation invariant feature [56, 58, 12], and pose normalization [59] methods have been developed and can be used to subdivide highly heterogeneous subtomograms through unsupervised clustering. However, these approaches do not take into account of the missing wedge effect, which introduces anisotropic resolution and is not rotation invariant. In addition, such approaches have limited structural discrimination ability in the presence of high level of noise in the subtomograms.

We aim to overcome the aforementioned challenges and limitations of structural mining in cellular tomograms by complementing with existing approaches. In this paper, we propose to use supervised deep learning approach to subdivide a large number of structurally highly heterogeneous subtomograms into structurally more homogeneous smaller subsets with significantly improved accuracy and scalability. After the subdivision, the computationally intensive reference-free structural recovery approaches can be separately applied to selected subsets in a divide and conquer fashion, which would significantly reduce the overall computation cost.

The major component of our new approach is a Convolutional Neural Network (CNN) classifier. Due to its superior scalability and good generalization ability, CNNs have made it computationally feasible to use a large number of parameters to approximate the complex mapping from massive data to class labels. In this paper, We propose tailored 3D variants (Section 2.2) of two popular CNN image classification models. These two CNN models have achieved state-of-the-art supervised classification accuracy on popular image classification benchmark datasets (e.g., ImageNet Dataset [45]). The first model (Section 2.2.1) is characterized by relatively low depth and relatively complex parallel local filter structure (i.e. inception structure [51]). The second model (Section 2.2.2) is characterized by relatively high depth and very small simple convolution filters. In addition, because the inputs of the models are 3D gray-scale images (i.e. subtomograms) representing the 3D structures of particles contained in the image, it is important for our CNN models to isotropically capture the inherent 3D spatial structure in such 3D images. Therefore, in our models we use single channel 3D filters for convolution and pooling, instead of the 2D filters used in common deep learning based computer vision applications.

The above CNN models are designed for supervised classification. Since the native structures of most macromolecular complexes are unknown [25, 58], there is a particular need for discovering macromolecular complex structures that *do not exist* in the training data. To do this, we combine CNN with unsupervised clustering (Section 2.3). First, we adapt the output layer of a trained CNN classifier to extract structure features that are invariant to both rigid transforms and missing wedge effect. Such structural feature extraction is equivalent to performing a *non-linear projection* of the testing subtomograms to the structural space *spanned* by the structures in the training data, an analogy to distance metric learning [54]. Then, we subdivide the projected subtomograms using unsupervised clustering, and recover the structures independently using reference-free classification and averaging [55, 20].

Our experiments on realistically simulated subtomograms show that the deep structural features extracted by the our CNN models are significantly faster and more robust to imaging noise and missing wedge effect than our previously used rotation invariant features [56, 58] approach. Clustering in the deep structural feature space produced significantly more evenly distributed clusters than our previous approach of clustering of pose normalized subtomograms [59]. Our proof-of-principle experiments on experimental subtomograms of purified macromolecular complexes also achieved competitive classification performance. Therefore, our experiments validate that our deep learning based approach is in practice a significantly better choice for subdividing millions of subtomograms. More importantly, our experiments (Section 3.3) on simulated data demonstrate that our approach is able to recover new structures that do not exist in the training data.

## 2 Methods

### 2.1 Background

In recent years, deep learning has emerged as a powerful tool for many computer vision tasks, such as image classification and object detection. Deep learning has achieved state-of-the-art supervised image classification performance on popular benchmark image datasets such as ImageNet [45], which contains more than 14 million images separated into at least 1000 classes. The CNN [32] is one of the most important techniques in deep learning. It is composed of alternating convolutional layers and pooling layers, and one or more fully connected layers. By utilizing multiple stacked processing layers to represent features of data, it allows learning increasingly abstract features at increasing scales.

More precisely, each convolutional layer consists of a set of learnable filters in the form of neurons with shared weights. Each neuron in this layer is connected to a region of neighbouring neurons in the previous layer. The outcome of the convolutional layer is obtained by first convolving with the output of the previous layer with learned filters and then applying a nonlinear activation function, such as sigmoid, tanh or a rectified linear unit (ReLU) [23], on the convolved results. The pooling layer is a form of down sampling used to reduce computation cost. Calculating the local maximum (max pooling) or average (average pooling) values are common forms of such pooling. After several rounds of convolutional and pooling layers, one or more fully connected layers are usually added to extract more global features. As the name suggests, each unit in these layers connects to all units from the previous layer. The last output layer is usually a softmax layer for multi-class classification tasks, generating a probability distribution over all possible classes. Other classifiers can also be used on those extracted features by CNN.

In 2012, the CNN architecture AlexNet proposed by Krizhevsky et al [30], first showed significant performance improvements on the supervised image classification tasks compared to the traditional methods. Since then, CNN has became the dominant approach for large scale supervised image classification tasks, and more advanced architectures have been developed, such as GoogleNet (inception network) [50], VGGNet [48], and ResNet [26]. Here, we developed two 3D CNN models for supervised subtomogram classification and feature extraction.

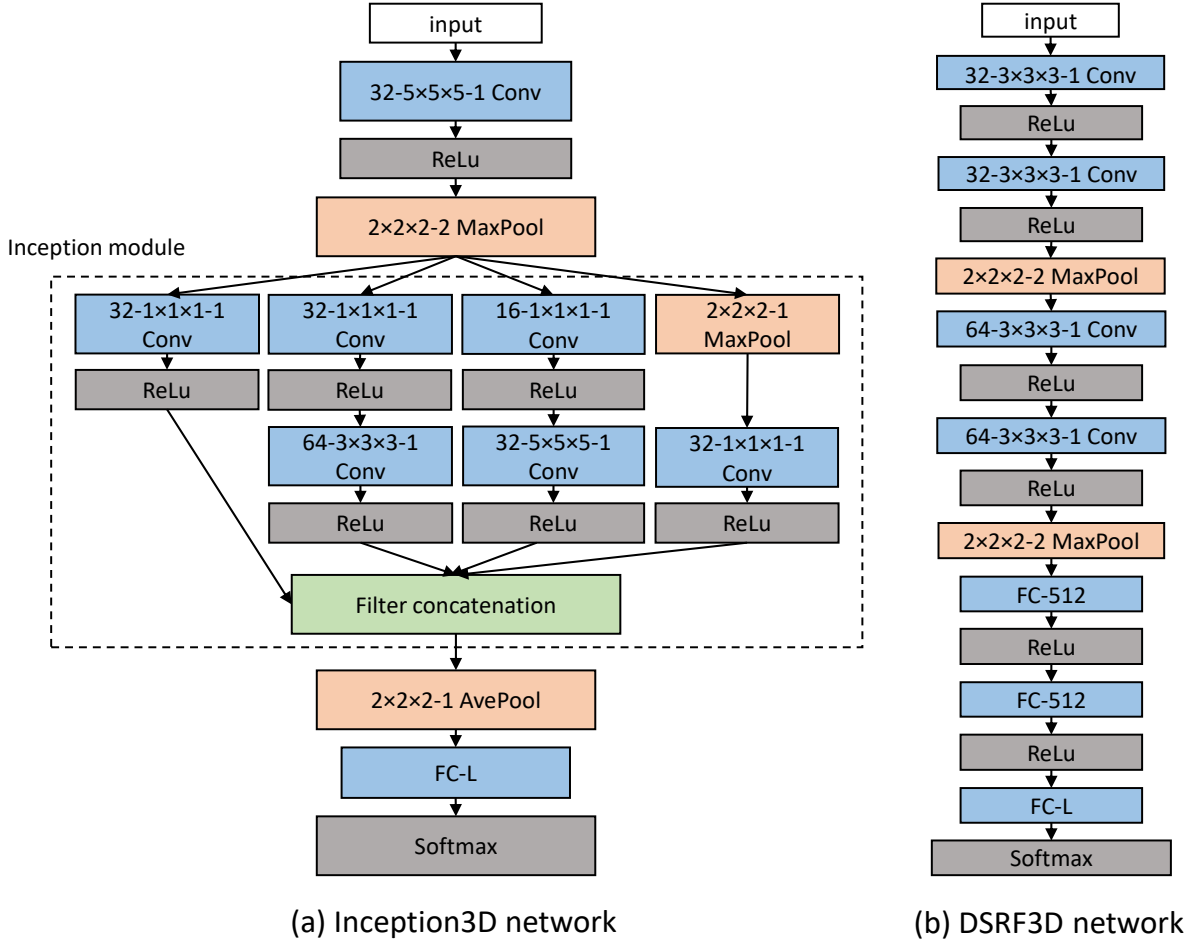

Figure 1: Architectures of our convolutional neural networks. (a) Inception3D network. (b) DSRF3D network. These networks both stack multiple layers. Each box represents a layer in the network. The type and configuration of layer are listed in each box. For example, "32-5x5x5-1 Conv" denotes a 3D convolutional layer with 32 5x5x5 filters and stride 1. "2x2x2-2 MaxPool" denotes a 3D max pooling layer implementing max operation over 2x2x2 regions with stride 2. 'FC-512' and 'FC-L' denote a fully connected linear layer with 512 and L neuron units respectively, where every unit is connected to every output of the previous layer. L is the number of classes in the training dataset. 'ReLU' and 'Softmax' denote different types of activation layers.

## 2.2 Convolutional neural network based supervised subtomogram classification

### 2.2.1 Inception3D network

In this section, we propose a 3D variant of tailored inception network [50], denoted as Inception3D. Inception network has the ability to achieve competitive performance with relatively low computational cost [50]. The architecture of our model is shown in Figure 1a. It contains one inception module [50], where 1x1x1, 3x3x3, and 5x5x5 3D filters are combined with 2x2x2 3D max pooling layer. The filters are implemented in parallel and concatenated, so that the features extracted at multiple scales using filters of different sizes are simultaneously presented to the following layer. The 1x1x1 filters before the 3x3x3 and 5x5x5 convolutions are designed for dimensional reduction. The inception module is followed by a 2x2x2 average pooling layer, then by a fully connected output layer with the number of units equal to the structure class number. All hidden layers are equipped with the rectified linear (ReLU) activation. The output is a fully connected layer with a softmax activation layer.

### 2.2.2 DSRF3D network

In this section, we propose a 3D variant of tailored VGG network (VGGNet) [48], denoted as Deep Small Receptive Field (a.k.a DSRF3D). The architecture of out model is shown in Figure 1b. Compared with the Inception3D network, DSRF3D is featured with deeper layers and very small 3D convolution filters of size 3x3x3. The stacking of multiple small filters has the same effect of one large filter, with the advantages of less parameters to train, and more non-linearity [48]. The architecture consists of four 3x3x3 3D convolutional layers and two 2x2x2 3D max pooling layers, followed by two fully connected layers, then followed by a fully connected output layer with the number of units equal to the structure class number. All hidden layers are equipped with the ReLU activation layers. The output is a fully connected layer with a softmax activation layer.

## 2.3 Combination of supervised structural feature extraction and unsupervised clustering for structural discovery

For the multi-class classification tasks in Section 2.2, the last fully connected layers' activation function used in Sections 2.2.1 and 2.2.2 are softmax functions:

$$o_j^{\text{softmax}}(\mathbf{x}) = P(j|\mathbf{x}) = \frac{e^{f_j(\mathbf{x})}}{\sum_{l=1}^{L} e^{f_l(\mathbf{x})}} \tag{1}$$

, where

$$f_j(\mathbf{x}) = \mathbf{x}^\top \mathbf{w}_j \tag{2}$$

, $\mathbf{x}$ are the inputs of the last fully connected layer, $\mathbf{w}_j$ are the weights associated with the $j$th class, $f_j(\mathbf{x})$ is the output of the last fully connected layer associated with the $j$th class, and $P(j|\mathbf{x})$ is the probability of the subtomogram is assigned to class $j$.

Designed for multi-class classification, the softmax activation $o_j^{\text{softmax}}$ re-scales $f_j$ exponentially. Therefore, it encourages output towards binary values, which reduces the extracted structural feature information that are useful for precisely subdividing input subtomograms. Once a CNN is trained for the classification task, we remove the softmax activation layer to obtain the linear activation of the last fully connected layer:

$$o_j^{\text{linear}}(\mathbf{x}) = f_j(\mathbf{x}) \tag{3}$$

. Using linear activation, we obtain a more continuous representation of the tendency that a subtomogram is predicted to belong to a class. Such continuous outputs produce structural features that are invariant to rigid transformation and missing wedge effect, representing a *nonlinear projection* of a subtomogram to a low dimension space *spanned* by structural classes in the training data. In principle, such features can also be extracted from hidden layers, providing richer structural descriptions, as long as they are invariant to rigid transformation of the particle, and invariant to missing wedge effect.

To apply such deep structural feature extraction to the discovery of new structures, after the above dimension reduction, we over-partition the subtomograms using k-means clustering to obtain a finer subdivision of subtomograms. The unsupervised reference free classification [e.g. 55, 5, 47] (or structural pattern mining [59]) is then independently applied to each cluster of subtomograms to recover the representative structures in the cluster.

### 2.3.1 Implementation details

The network training and testing code is implemented using Keras [14] and Tensorflow [1]. The Keras_extras library [29] is used for multiple GPU parallelization. A variant of our Tomominer library [20, 59] is used for reference-free subtomogram classification and other processing. The experiments are performed on a computer equipped with two Nvidia GTX 1080 GPUs, one Intel Core i7-6800K CPU, and 128GB memory.

For the baseline method, the calculation of rotation invariant features is based on SHTools [52]. K-means clustering and Support Vector Machine (SVM) based supervised multi-class classification are performed using the Sklearn toolbox [42].

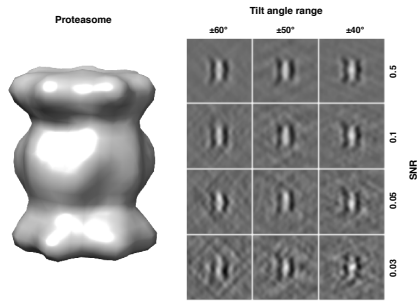

Figure 2: Left: isosurface of density map of yeast 20S proteasome (PDB ID: 3DY4). Right: Center slices (in parallel with x-z plane) in the simulated tomograms with different degree of SNRs and tilt angle ranges.

## 3 Results

### 3.1 Datasets generation

#### 3.1.1 Simulated subtomograms from known structures

For a reliable assessment of the approaches, we generated subtomograms by simulating the actual tomographic image reconstruction process in a similar way as previous works [16, 6, 40, 57], with the proper inclusion of noise, and missing wedge effect, and electron optical factors, such as the Contrast Transfer Function (CTF) and Modulation Transfer Function (MTF). Specifically, macromolecular complexes have an electron optical density proportional to the electrostatic potential. We used the PDB2VOL program from the Situs [53] package to generate volumes of $40^3$ voxels with a resolution and voxel spacing of 0.92nm. The density maps are used to simulate electron micrograph images through a set of tilt-angles. For this paper we set typical tilt-angle ranges of $\pm 60°$, $\pm 50°$ and $\pm 40°$. We added noise to electron micrograph images [16] to achieve the desired SNR levels, whose range cover the SNRs estimated from experimental data (Section 3.1.2). Next we convoluted the electron micrograph images the CTF and MTF to simulate optical effects [18, 40]. The acquisition parameters used are typical of those found in experimental tomograms [60], with spherical aberration of 2 mm, defocus of -5$\mu$m, and voltage of 300kV. The MTF is defined as $\text{sinc}(\pi\omega/2)$ where $\omega$ is the fraction of the Nyquist frequency, corresponding to a realistic detector [37]. Finally a direct Fourier inversion reconstruction algorithm (implemented in the EMAN2 library) [21] is used to produce the simulated subtomogram from the tilt series. Figure 2 shows examples of such simulated subtomograms with different SNR.

We collected 22 macromolecular complexes from the Protein Databank (PDB) [7] (Table 1). We constructed a simulated dataset for each pair of SNR and tilt angle range parameters. Inside a dataset, for each complex, we generated 1000 simulated subtomograms that contain randomly rotated and translated particle of that complex. Furthermore, we also simulated 1000 subtomograms that contain no particle. As an outcome, dataset contains 23,000 simulated subtomograms of 23 structural classes.

#### 3.1.2 CryoEM data collection, tomogram reconstruction, and preparation of ground truth

We captured tomograms of purified *E. coli* Ribosome and human 20S Proteasome through similar procedure as [60]. Specifically, Cryo-Electron Microscopy was performed at 300 keV using a TF30 "Polara" electron microscope (FEI) equipped with a Quantum postcolumn energy filter (Gatan) operated in zero-loss imaging mode with a 20-eV energy-selecting slit. Images were recorded on a postfilter $\approx 4,000 \times 4,000$ K2-summit direct electron detector (Gatan) operated in counting mode with dose fractionation, with a calibrated pixel size of 0.23nm at the specimen level. Tilt series were collected using SerialEM [36] at defocus ranges of -6 to -5$\mu$m. During data collection, the autofocusing routine was iterated to achieve a very stable defocus through the tilt series with 100nm accuracy. Tomographic reconstructions were calculated in IMOD program using weighted back-projection [46]. The reconstructed tomograms were then four times binned to a voxel spacing of 0.92nm.

To prepare for ground truth, we performed template-free particle picking similar to [43] through convoluting the tomograms with 3D Difference of Gaussian function with scaling factor of $\sigma = 7$nm and scaling factor ratio $K = 1.1$ to extract an initial set of 3,646 subtomograms of size $40^3$ voxels. The extracted subtomograms were smoothed by convoluting with a Gaussian kernel of $\sigma = 1$nm. We then aligned the subtomograms against Proteasome and Ribosome templates. These templates were obtained from first generating 4nm resolution density maps from the PDB structures using PDB2VOL program [53], then convoluting the density maps with proper CTF according to experimental data (Section 3.1.2). The subtomograms with high alignment scores were selected. Finally, a set of 401 subtomograms were obtained, 201 and 200 were labeled as Proteasome and Ribosome respectively.

To estimate SNR, for each structural class, we randomly selected 100 pairs of subtomograms that were aligned with the corresponding template, and estimated the SNR given each subtomogram pair according to [19]. The mean SNRs are 0.06 and 0.08 for Proteasome and Ribosome respectively.

| PDB ID | Macromolecular complex |
|--------|------------------------|
| 1A1S | Ornithine carbamoyltransferase |
| 1BXR | Carbamoyl phosphate synthetase |
| 1EQR | Aspartyl tRNA-synthetase |
| 1F1B | E. coli aspartate transcarbamoylase |
| 1FNT | Yeast 20S proteasome with activator |
| 1GYT | Aminopeptidase a |
| 1KP8 | GroEL-KMgATP 14 |
| 1LB3 | Mouse L-chain ferritin |
| 1QO1 | Rotary motor in ATP synthase |
| 1VPX | Transaldolase |
| 1VRG | Propionyl-CoA carboxylase, beta subunit |
| 1W6T | Octameric enolase |
| 1YG6 | ClpP |
| 2BO9 | Human carboxypeptidase A4 |
| 2BYU | Small heat shock protein Acr1 |
| 2GHO | Recombinant thermus aquaticus RNA polymerase |
| 2GLS | Glutamine synthetase |
| 2H12 | Acetobacter aceti citrate synthase |
| 2IDB | 3-octaprenyl-4-hydroxybenzoate decarboxylase |
| 2REC | RecA hexamer |
| 3DY4 | Yeast 20S proteasome |
| 4V4Q | Bacterial ribosome |
| NULL | (No particle) |

Table 1: Macromolecular complexes as structural classes used for simulation study.

| SNR / Tilt angle range | ±60° | | | ±50° | | | ±40° | | |
|---|---|---|---|---|---|---|---|---|---|
| | Inception3D | DSRF3D | RIF-SVM | Inception3D | DSRF3D | RIF-SVM | Inception3D | DSRF3D | RIF-SVM |
| 1000 | 0.993 | 0.990 | 0.992 | 0.994 | 0.978 | 0.983 | 0.983 | 0.991 | 0.967 |
| 0.5 | 0.975 | 0.972 | 0.929 | 0.964 | 0.967 | 0.885 | 0.931 | 0.951 | 0.857 |
| 0.1 | 0.851 | 0.891 | 0.762 | 0.807 | 0.873 | 0.633 | 0.809 | 0.866 | 0.649 |
| 0.05 | 0.757 | 0.767 | 0.592 | 0.682 | 0.728 | 0.455 | 0.637 | 0.684 | 0.468 |
| 0.03 | 0.608 | 0.658 | 0.446 | 0.516 | 0.604 | 0.319 | 0.473 | 0.556 | 0.341 |

Table 2: The classification accuracy of simulated datasets of subtomograms at different levels of SNR and tilt angle range.

## 3.2 Classification performance

### 3.2.1 On simulated data

To assess the classification performance, for each dataset generated in Section 3.1.1, we randomly separated the subtomograms into two equal sized sets. We used one set for training, and the other set for testing.

The CNN models were trained using stochastic gradient descent (SGD) with Nesterov momentum of 0.9 to minimize the categorical cross-entropy cost function. The initial learning rate was set to 0.01, with a decay factor of 1e-6. A 70% dropout [49] was implemented in Inception3D network to prevent over-fitting, i.e., a unit in network was retained with probability 70% during the training. SGD training were performed with a batch size of 64 for 20 epochs.

For the baseline method, we used spherical harmonics rotation invariant feature [e.g. 58, 56] in combination with Support Vector Machine (SVM) with Radial Basis function kernel, denoted as RIF-SVM.

The classification accuracy is summarized in the Table 2. It can be seen that, at realistic SNR and tilt angle range levels, all CNN models achieved significantly higher classification accuracy than the rotation invariant feature based method.

We further measured the computation speed. On average, the training time took 0.0034 and 0.0055 seconds per subtomogram per epoch for Inception3D and DSRF3D networks respectively. Given trained models, the feature extraction and classification take 0.0015 and 0.0017 seconds per subtomogram for Inception3D and DSRF3D networks respectively. Thus, after training, CNN based feature extraction and classification of one million subtomograms would take less than one hour on a single affordable desktop computer with two affordable GPUs. By contrast, the unsupervised rotation invariant feature extraction took 0.1 second per subtomogram. With fixed subtomogram size and class number, the CNN models scales linearly respect to the number of subtomograms. By contrast, SVM scales quadratically respect to the number of subtomograms.

### 3.2.2 On experimental data

We randomly split the subtomograms into two equal sized sets and used one set as training and the other set as testing. In the training set, 105 and 95 subtomograms are labeled as Proteasome and Ribosome respectively. In the testing set, 96 and 105 subtomograms are labeled as Proteasome and Ribosome respectively.

Although the number of samples was significantly smaller than the typical sample size in for deep learning tasks, the Inception3D network still achieved a classification accuracy of 0.905, which is higher than the classification accuracy of 0.890 of the baseline method of rotation invariant feature in combination with SVM. DSRF3D network fail to converge during training due to small sample size.

## 3.3 Detecting new structures

In this section, we test if our approach in Section 2.3 can be used to facilitate the recovery of structures that do not exist in the training data. The experiments were performed using subtomograms simulated at SNR 0.05 and tilt angle range ±60° (Section 3.1.1).

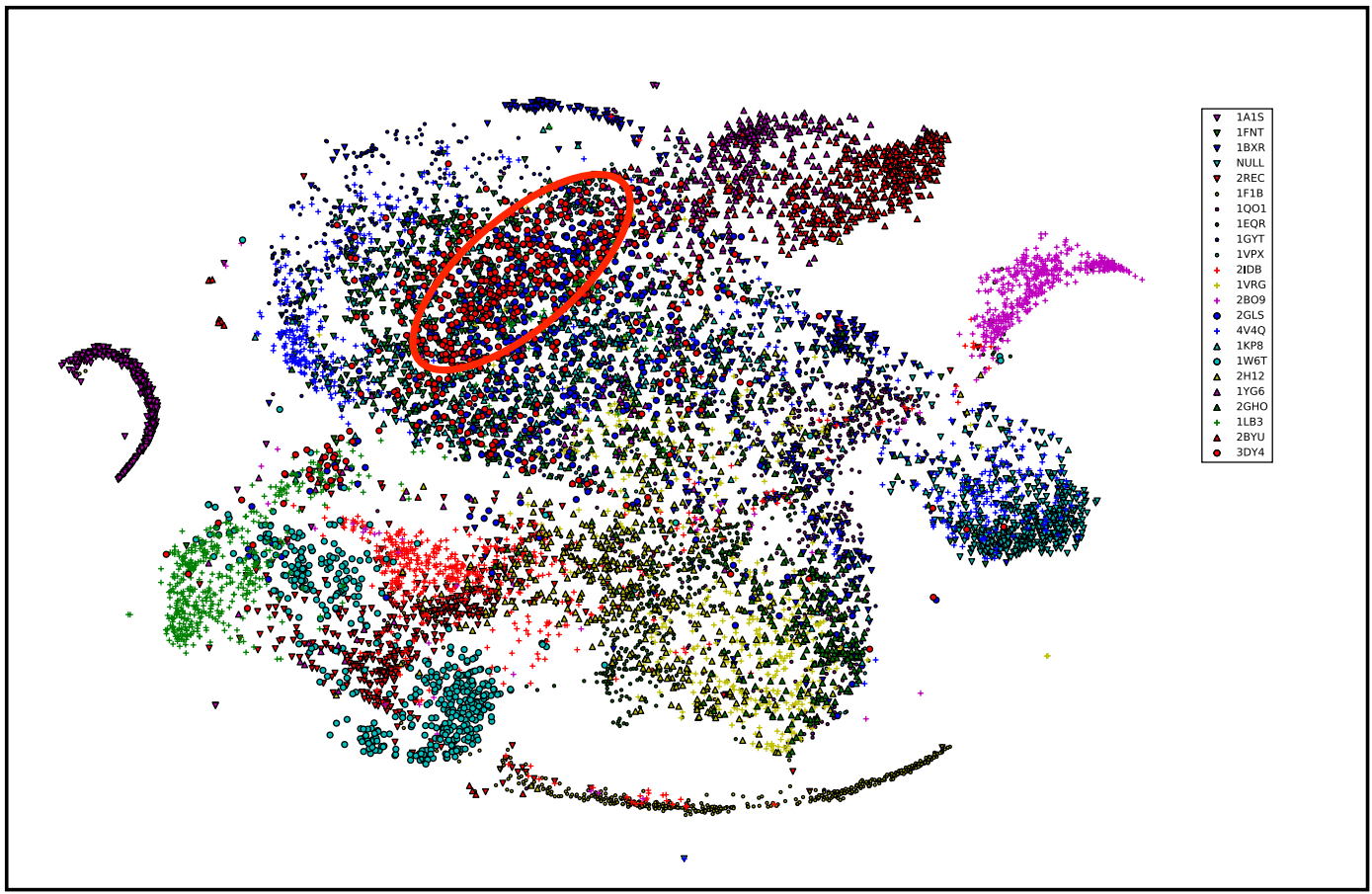

Figure 3: Subtomograms in the test set projected to the structural feature space of $\mathbb{R}^{22}$ through structural feature extraction (Section 2.3). The projected subtomograms were further embedded to $\mathbb{R}^2$ using T-SNE [35] only for visual inspection. The points were shaped and colored according to their true class labels. The region enriched with Proteasome subtomograms (PDB ID: 3DY4) was highlighted using red circle.

We prepared a training set $S^{\text{train}}$ with all 23 structural classes except Proteasome (PDB ID: 3DY4), and a test set $S^{\text{test}}$ with all 23 structural classes. There are 500 subtomograms in each class in each set. We trained an Inception3D network using $S^{\text{train}}$, then used the trained network to extract the structural features by projecting the subtomograms of $S^{\text{test}}$ into a 22 dimensional deep structural feature space $\mathbb{R}^{22}$ corresponding to the 22 classes in the training data. In such case, each subtomogram in $S^{\text{test}}$ correspond to one point in $\mathbb{R}^{22}$. For visual inspection, we further embedded the points in $\mathbb{R}^{22}$ into a two dimensional space $\mathbb{R}^2$ using the T-SNE algorithm [35], which is particularly well-suited for embedding high-dimensional data into a space of two or three dimensions for visualization. Figure 3 shows the embedded points. It is evident that samples are generally concentrated in subregions according to their structural classes. Most importantly, although Proteasome subtomograms do not exist in the $S^{\text{train}}$, the Proteasome subtomograms in $S^{\text{test}}$ are still concentrated at certain subregion in $\mathbb{R}^2$ (Figure 3).

Inspired by the above observations, we systematically examined the possibility of recovering new structures using our approach (Section 2.3) by conducting leave-one-out test to all 22 macromolecular complex structure classes. For each test, we removed subtomograms of a class $C^{\text{true}}$ from training data, then trained an Inception3D network, we then used the trained network to project the subtomograms of $S^{\text{test}}$ into the deep structural feature space $\mathbb{R}^{22}$ according to Section 2.3. Then we performed k-means clustering in $\mathbb{R}^{22}$ with k=100 clusters. We found the cluster $L^{\text{pred}}$ in which particles of $C^{\text{true}}$ were most enriched. We then applied unsupervised reference-free subtomogram classification and averaging [55, 20] (with 3 classes, 5 iterations) to the subtomograms of $L^{\text{pred}}$. Among the classes predicted by unsupervised subtomogram classification, we identified the class $C^{\text{pred}}$ that was mostly enriched with particles of $C^{\text{true}}$. We then calculated the structural discrepancy between the subtomogram average of $C^{\text{pred}}$ and the true structure of $C^{\text{true}}$. Such *structural discrepancy* is measured using Fourier Shell Correlation (FSC) [33] with 0.5 cutoff, representing the maximal size of the structural factors that are discrepant between two structures. When using a structural discrepancy of 7nm to determine whether the structure prediction is success, we found 16 out of the 22 leave one out tests correctly recovered structures of $C^{\text{true}}$ that (Figure 4) even $C^{\text{true}}$ does not exist in $S^{\text{train}}$.

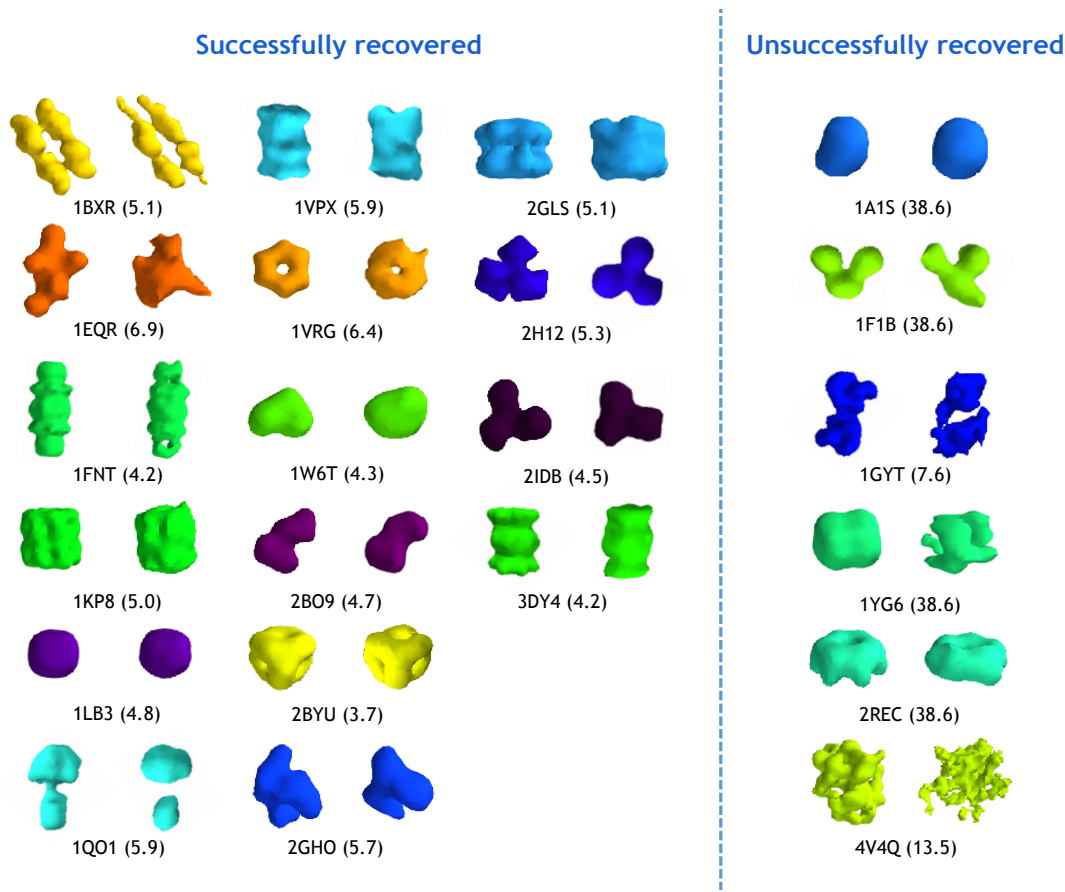

Figure 4: The isosurfaces of true (left) and predicted (right) structures. The predicted structures were obtained by our approach (Section 2.3). The numbers in parentheses were structural discrepancy between true and predicted structures.

We further performed the same test using DSRF3D network, and we were able to get similar results. Specifically, 18 out of 22 structures were recovered (Table 3).

We further inspected the cluster size distribution of the result of k-means clustering of subtomograms projected to the feature space $\mathbb{R}^{22}$ used for Figure 3. The cluster sizes did not vary too much (Figure 5a). By contrast, when applying our previous pose-normalization method [59] to subtomograms in $S^{\text{test}}$, then perform k-means clustering on the pose normalized subtomograms to subdivide the 11,500 subtomograms in the $S^{\text{test}}$ into 100 clusters, we found that most clusters are very small (Figure 5b). Specifically, there were 84 clusters whose size $\leq 10$. On the other hand, there were 4 large clusters with size $> 1000$, covering 6089 subtomograms, with mixed particles of similar structural sizes. The largest cluster had a size of 2470. The highly uneven cluster size distribution was due to the reduced discrimination ability of distance matrics in high dimensional space [2]. Compared with our previously used pose-normalization approach, our supervised deep structural feature extraction approaches achieved significantly better subtomogram subdivision ability.

## 4 Discussion

CECT is currently the preferred experimental tool to visualize macromolecular complexes in near native conditions at submolecular resolution, when coupled with deep data mining it emerges as a very promising tool for systematic detection of structures and spatial organizations inside single cells. However, due to high level of structural complexity and practical imaging limitations, systematic *de novo* structural discovery of macromolecules from such tomograms requires the computational analysis of large amount of subtomograms. Existing structural recovery approaches are through reference-free subtomogram averaging [10], classification [e.g. 55], or structural pattern mining [59], and they have very limited scalability. Therefore, efficient and accurate subdivision of large amount of highly heterogeneous subtomograms is a key step for scaling up such computational intensive structural recovery approaches. On the other hand, our previously used rotation invariant feature [56, 58] and pose normalization [59] subdivision approaches have limited discrimination ability and scalability. To complement existing approaches, as a proof-of-principle, in this work we propose to use deep learning based supervised approaches to significantly improve both scalability and discrimination ability of subtomogram subdivision. Our preliminary results demonstrated superior performance over our previously used subdivision approaches [56, 58, 59], and the potential of using our new approach for the discovery of new structures.

| PDB ID | Structural Discrepancy | PDB ID | Structural Discrepancy |
|---|---|---|---|
| 1A1S | 4.3 | 1W6T | 38.6 |
| 1BXR | 5.5 | 1YG6 | 4.5 |
| 1EQR | 5.7 | 2BO9 | 4.7 |
| 1F1B | 5.1 | 2BYU | 5.3 |
| 1FNT | 4.6 | 2GHO | 6.5 |
| 1GYT | 9.9 | 2GLS | 3.6 |
| 1KP8 | 4.7 | 2H12 | 4.9 |
| 1LB3 | 5.5 | 2IDB | 4.6 |
| 1QO1 | 4.3 | 2REC | 38.6 |
| 1VPX | 4.2 | 3DY4 | 4.4 |
| 1VRG | 5.0 | 4V4Q | 9.4 |

Table 3: Structural discrepancy (in nm unit) between true structures and structures predicted using our approach (Section 2.3) with DSRF3D network

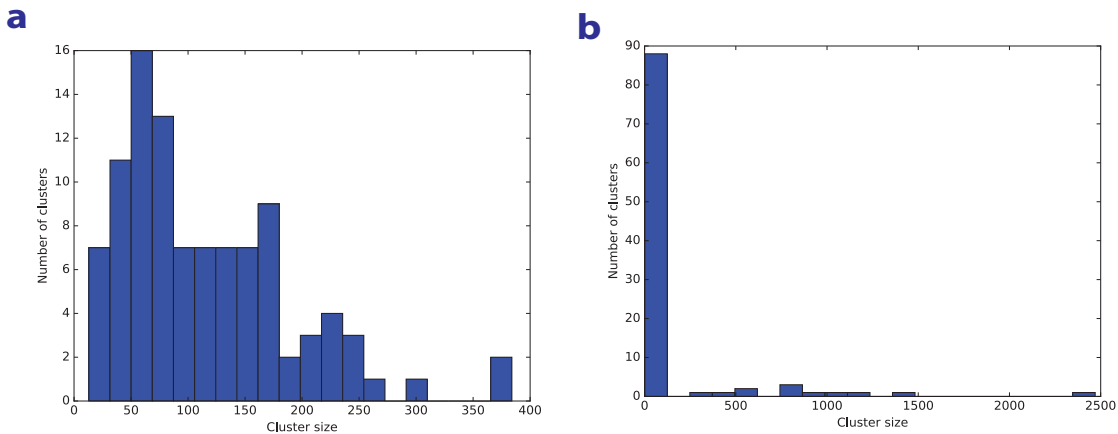

Figure 5: Cluster size distribution of kmeans clustering of (a) CNN extracted structural features and (b) pose-normalized subtomograms.

Potential uses of our approach are to quickly subdivide the highly heterogeneous particles into subsets, and separately recover the representative structures in each selected subset using computation intensive unsupervised subtomogram classification or pattern mining approaches. Given a recovered structure, one can further verify whether it already exist in training data. The particles of the new structures can be further included into training data for more comprehensive disentangling of structural features with enhanced discrimination ability. This work represents an important step towards large scale systematic detection of structures and spatial organizations of large macromolecular complexes inside single cells captured by CECT data. Our approach can also be applied to similar analysis tasks arisen in cryo tomograms of cell lysate or purified complexes. As our approach involves supervised training, our method relies on the availability and quality of training data. In practice, the training data can come from diverse sources. They can be from cryo tomograms of purified complexes captured in the same imaging condition as test samples. They can also be from particles in CECT images located through different approaches, such as correlated super-resolution imaging [11, 28], template search [6, 31], unsupervised reference-free subtomogram classification [e.g. 55], or structural pattern mining [59]. On the other hand, the proper strategies of constructing and processing training data remain to be explored. In addition, the proposed CNN architectures remain to be further optimized for improved performance.

# 5 Acknowledgment

We thank Dr. Robert Murphy for providing invaluable suggestions. We also thank Mr. Kshitiz Dange for providing technical support.

*Funding*: This work was supported in part by U.S. National Institutes of Health (NIH) grant P41 GM103712. XC acknowledges support of a Ji-Dian Liang Graduate Research Fellowship. TZ acknowledges support of Sir Henry Dale Fellowship jointly funded by the Wellcome Trust and the Royal Society (Grant Number 107578/Z/15/Z). GY acknowledges support of National

Science Foundation (NSF) Faculty CAREER Grant DBI-1149494. EX acknowledges supports of NIH Grant R01GM114311, NIH Grant P30DA035778, and Pennsylvania Department of Health CURE Grant BD4BH4100070287.

*Conflict of Interest*: none declared.